# Stable mode-locked pulses from mid-infrared semiconductor lasers

Christine Y. Wang[1], L. Kuznetsova[2], V. M. Gkortsas[3], L. Diehl[2] , F. X. Kärtner[3], M. A. Belkin[2], A. Belyanin[4], X. Li[2], D. Ham[2], H Schneider[5], P. Grant[6], C. Y. Song[6], S. Haffouz[6], Z. R. Wasilewski[6], H. C. Liu[6] & Federico Capasso[2]

*[1]Department of Physics and [2]School of Engineering and Applied Sciences, Harvard University, Cambridge, MA 02138, USA*

*[3]Research Laboratory of Electronics, Massachusetts Institute of Technology, Cambridge, MA 02139, USA*

*[4]Department of Physics, Texas A&M University, College Station, Texas 77843, USA*

*[5]Institute of Ion Beam Physics and Materials Research, Forschungszentrum Dresden-Rossendorf, 01314, Dresden, Germany*

*[6]Institute for Microstructural Sciences, National Research Council, Ottawa K1A 0R6, Ontario, Canada*

**Stable trains of ultrashort light pulses with large instantaneous intensities from mode-locked lasers are key elements for many important applications such as nonlinear frequency conversion [1-3], time-resolved measurements [4, 5], coherent control [6, 7], and frequency combs [8]. To date, the most common approach to generate short pulses in the mid-infrared (3.5-20 µm) molecular "fingerprint" region relies on the down-conversion of short-wavelength mode-locked lasers through nonlinear processes, such as optical parametric generation [9-11] and four-wave mixing [12]. These systems are usually bulky, expensive and typically require a complicated optical arrangement. Here we report the unequivocal demonstration of mid-infrared mode-locked pulses from a semiconductor laser.**

**The train of short pulses was generated by actively modulating the current and hence the optical gain in a small section of an edge-emitting quantum cascade laser (QCL). Pulses with pulse duration at full-width-at-half-maximum of about 3 ps and energy of 0.5 pJ were characterized using a second-order interferometric autocorrelation technique based on a nonlinear quantum well infrared photodetector. The mode-locking dynamics in the QCLs was modelled and simulated based on Maxwell-Bloch equations in an open two-level system. We anticipate our results to be a significant step toward a compact, electrically-pumped source generating ultrashort light pulses in the mid-infrared and terahertz spectral ranges.**

Quantum cascade lasers (QCLs) [13], since their invention in 1994, have become the most prominent coherent light sources in the mid-infrared. One of the most striking differences between these unipolar devices and diode lasers is that their emission wavelength, gain spectrum [14], carrier transport characteristics, and optical dispersion can be engineered. This remarkable design freedom makes QCLs a unique candidate to serve as a semiconductor source of ultra-short pulses in the mid-infrared.

There is, however, an obstacle of fundamental origin that has so far prevented achieving ultrashort pulse generation in QCLs [15, 16]. In intersubband transitions, the carrier relaxation is extremely fast because of optical phonon scattering. As a result, the gain recovery time in QCLs is typically on the order of a few picoseconds [17, 18] which is an order of magnitude smaller than the cavity roundtrip time of 40-60 ps for a typical 2-3mm-long laser cavity. According to conventional mode-locking theory, this situation prevents the occurrence of stable passive mode-locking and impedes the formation of high-intensity pulses through active mode-locking [19].

In this work, we achieve stable mode-locking by designing a QCL structure with longer gain recovery time than conventional QCL designs and actively modulating the pumping current of a small section at one end of the laser cavity to provide net gain to the pulse. To our knowledge, this is also the first stable mode-locking of a laser with fast gain recovery time. Our QCL structure, shown in Figure 1a, is based on a "diagonal transition" in real space [20], i.e. the laser transition takes place between levels confined in two adjacent wells separated by a thick barrier. The reduced wavefunction overlap between the upper and lower laser states results in a phonon-limited upper state lifetime of up to 50 ps below threshold. As we will see later, we are able to identify a parameter range close to the laser threshold where we can form stable mode-locked pulses by modulating the gain in the small section of a device. High above threshold, pulses are broadened by the increased gain saturation in the un-modulated long section, and finally become unstable due to the increased spatial hole-burning and accelerated gain recovery.

There were several reports of mode-locking in QCLs [21-23], whose evidence were based on broadband optical spectra with a large number of longitudinal modes and a narrow microwave beat note in the power spectrum at the laser roundtrip frequency, which indicated that the electric field waveform circulating in the laser cavity and thus the phase relationship between the longitudinal modes was stable over a large number of round-trips. However, due to the lack of a suitable apparatus for second-order autocorrelation measurements, no direct evidence was given to demonstrate that the circulating waveform was indeed a periodic sequence of isolated pulses, which would result from all modes having equal and stable phases. Subsequent pulse characterization using autocorrelation techniques showed that under the previous conditions, the output of free running QCLs was not composed of one isolated pulse per roundtrip [15, 16]. The physics of multimode behaviour observed in those lasers is described by spatial hole-burning and the Risken-Nummedal-Graham-Haken (RNGH) instability [15, 16], rather than by stable mode locking.

Our devices were processed into ridge waveguides with multiple electrically independent sections, see Fig. 1b and a processing description in Methods. Radio-frequency (RF) signals were injected into a short 120 μm to 160 μm-long section at one end of the 2.6 mm-long laser ridge via a bias tee to modulate the pumping current in the section. The current-voltage (I-V) and light-current (L-I) characteristics of a 16μm-wide-ridge device when the ridge is pumped in continuous wave with no RF modulation at 77K are shown in Figure 2a. The differential resistance of the RF section above threshold is $R \approx 30\ \Omega$. Given the estimated capacitance across the insulation layer $C \approx 0.05$ pF, the RC-limited frequency response extends to ~100 GHz, far above the roundtrip frequency of the laser cavity. The actual microwave power delivered to the device measured by a network analyzer was about 30% of the input power, which indicates a significant impedance mismatch between the source and the device. The largest contribution to the impedance mismatch is the inductance caused by the bond wire, estimated to be about 1 nH. (See Supplementary Information 1)

Figure 2b shows the optical spectra of the device, measured with a Fourier transform infrared spectrometer (FTIR), as a function of the RF modulation frequency, for a DC pumping current of 340 mA which is about 1.1 times the laser threshold. The RF input power was kept constant at 35 dBm. The laser emission is single mode when no gain modulation is applied. At a gain modulation frequency corresponding to the laser roundtrip frequency of 17.86 GHz we observe significant spectral broadening. The spectrum exhibits many longitudinal modes with an approximately Gaussian envelope. With a slightly detuned RF frequency from resonance, the laser spectrum becomes narrower. When the RF frequency is tuned further away from resonance (<17.36 GHz or >18.11 GHz), lasing stops. The RF power is able to suppress the lasing up to 355 mA of pumping current. As we shall see, stable mode-locked pulses are formed for pumping close to threshold; gain can overcome loss only if the RF frequency is tuned to resonance, and the laser generates a train of isolated short pulses which reach the modulated section at each round-trip at its gain maximum. When the laser is pumped at higher DC current (1.45 times the laser threshold), the optical spectra show similar resonance behaviour, but the RF modulation is no longer sufficiently strong to suppress the lasing when the detuning between the RF modulation frequency and the cavity round-trip frequency is large (Fig. 2c).

The laser emission is characterized using a second-order interferometric autocorrelation (IAC) technique [24] using a two-photon quantum well infrared photodetector (QWIP) [25, 26]. A description of the autocorrelator setup is given in Methods. For the case of stable, isolated periodic pulses, the ratio between the interference maximum and the background should be 8 to 1 [16]. Fig. 3a shows the IAC trace obtained when the QCL is pumped at 340 mA and the RF modulation is applied at the cavity resonance frequency of 17.86 GHz. The observed peak-to-background ratio is 8:1, indicating stable mode-locking and the circulation of an isolated pulse per roundtrip in the cavity. The pulse width, estimated from the width of the interference fringes, is approximately 3 ps. This value is approximately a factor of two larger than the width of a Fourier-transform limited pulse duration calculated from the optical spectrum in Fig. 1a. The estimated energy per pulse is close to 0.5 pJ. The observed IAC in Fig. 3a is in good agreement with a simulated IAC trace shown in Fig. 3d for the same ratio $p = 1.1$ of the pumping current to the threshold current.

When the DC pumping is increased to 450 mA, the peak-to-background ratio of the IAC is no longer 8:1, but rather 8:2.5, as shown in Fig. 3b. This indicates that the laser output does no longer consist of isolated pulses separated by the roundtrip time, but rather of overlapping pulses with a continuous-wave component. Further increasing the laser pumping current to 500 mA with fixed RF power decreases the peak-to-background ratio of the IAC even more to 8:4, see Fig. 3c.

We also measured the microwave spectrum of the laser output with applied RF modulation. The laser output was sent directly to a 20 GHz bandwidth QWIP [27] and the resulting photocurrent was displayed on a spectrum analyzer. The microwave spectrum of the laser output at DC pumping of 340 mA is shown in the inset of Fig. 3a. A narrow peak with a FWHM of ~3 kHz was observed at the RF modulation frequency, which indicates phase coherence between the modes for more than $10^6$ roundtrips. This beat note is also at least an order of magnitude narrower than any previously observed beat note from multimode QCLs [19, 23].

Figures 4a-c show the IAC for a 12 μm-wide laser with various RF power levels when the laser was pumped constantly at 265 mA (about 1.08 times the laser threshold). The pulse quality degrades dramatically as the RF input power decreases from 35 dBm to 12 dBm. Thus, operating near threshold and applying sufficiently large RF modulation are two necessary conditions to achieve stable mode-locking in our devices. For fixed input RF power (35 dBm) the IAC shows a similar behaviour to that of the 16 μm wide device as the DC pumping is increased.

Modulating the injection current of the QCLs leads to the modulation of the gain. In that case, the standard active mode-locking formalism of loss modulation which does not take the gain dynamics into account cannot be applied [18]. To understand the pulse regimes in our system, we use a simple model based on one-dimensional Maxwell-Bloch equations in a Fabry-Perot cavity, where the active medium is described by an open two-level system. This model has been shown to successfully describe the dynamics in QCLs without active modulation [15, 16]. Details of the equations and the simulation parameters can be found in the Supplemental Information 2.

In our model, we assume that the whole laser is DC pumped at $\lambda = p \times \lambda_{th}$, where $\lambda_{th}$ denotes the pumping at threshold, and p the pumping ratio. To model the active modulation in the short section, a sinusoidal term is added to the pumping: $\lambda = \lambda_{th} \times [p + m\sin(2\pi f_R t)]$, where *m* is the modulation amplitude, and $f_R$ is the cavity roundtrip frequency. Fig. 3d-f shows the simulated IACs for fixed modulation amplitude *m*=5 with different pumping ratio *p*, and Fig. 4d-f show the simulated IACs for fixed pumping ratio *p*=1.1 with different modulation amplitude *m*. For *p*=1.1 and *m*=5, the peak-to-background ratio of the IAC is 8:1, and the ratio gradually decreases with higher DC pumping *p* and smaller modulation amplitude *m*, which follows the experimental trend. Note however that there remain quantitative discrepancies with experiments, which show a faster decrease of the IAC ratio with increasing *p* and decreasing *m*. In order to predict the exact level to which the background in the IAC is rising, given certain pumping and modulation conditions, will need a much more detailed description of the gain as a function of pumping conditions and laser operation, then the two level model allows [18].

In both experiment and simulation, we observe side lobes in the IACs beside the main interference peak. The side lobes come from the multiple spikes in the intensity profile, as seen in the insets of Fig. 3d-f and Fig. 4d-f. Those spikes are caused by spatial hole-burning (SHB) in the laser gain created by standing waves in the cavity. The gain grating introduced by SHB interferes with active mode locking as it couples the longitudinal modes and reduces the phase locking imposed by the gain modulator. As a result, the modes develop nonlinear phases that lead to a waveform with multiple spikes. Note that in the experimental IACs, the 12 µm-wide device has more pronounced lobes than the 16 µm-wide one. This is consistent with our previous observation that SHB is less significant in wider devices because multiple transverse modes tend to wash out the gain grating formed by SHB [16].

In conclusion, we have demonstrated the first unequivocal stable mode-locking in QCLs via active gain modulation. As revealed by IAC, isolated periodic pulses as short as 3 ps can be generated in the vicinity of the laser threshold. Experimental data and numerical simulations show that the parameter window for stable mode-locking is determined both by the DC pumping current and the RF modulation power. Further pulse shortening and increase in pulse energy can be achieved by increasing gain lifetime and modulation amplitude.

## Methods

**Device.** The metalorganic chemical vapour deposition (MOCVD) growth started on a highly n-doped InP substrate ($n=3\times10^{18}$ $cm^{-3}$), with a 3 μm-thick InP waveguide cladding layer, n-doped to $n=1\times10^{17}$ $cm^{-3}$, and continued with the core consisting of a 0.3 μm-thick low n-doped ($n=1\times10^{17} cm^{-3}$) GaInAs layer followed by 35 stages of the active region and by a 0.3μm-thick GaInAs layer ($n=1\times10^{17} cm^{-3}$). The upper cladding was a 3.5 μm-thick InP layer, where the lower 3 μm was doped to $n=1\times10^{17}$ $cm^{-3}$, and the rest to $1\times10^{19}$ $cm^{-3}$. The topmost layer was composed of a 100 Å-thick InP layer followed by a 200 Å-thick GaInAs layer, both highly n-doped to $n=1\times10^{20}$ $cm^{-3}$ for plasmon enhanced confinement.

The material was processed into 2.6 mm long, 8-20 μm wide ridge waveguides by reactive ion etching. A 5 μm-thick layer of Microchem SU-8 2005 photoresist was used as an insulation layer between the wafer and the top contact to reduce the parasitic capacitance. The SU-8 2005 photoresist on top of the ridges was removed by standard photolithography and a metal layer (Ti/Au; 20 nm/300 nm) was then deposited to provide electrical contact. The top Ti/Au contact and the underlying heavily doped InP contact layers were etched out in specific areas along the ridges to create multiple electrically independent sections, see Fig. 1b, with minimal electrical crosstalk. A non-alloyed Ge/Au contact was deposited on the back. The samples were indium-soldered on copper holders and mounted in a liquid-nitrogen flow cryostat. The small section was aluminum-wire-bonded to a gold microstrip connected to an end launch connector and SMA cables for RF injection, while the rest of the ridge was bonded to a normal gold bonding pad.

**Measurements.** The optical power was measured by a thermal head power meter. The optical spectra were measured by a Nicolet 8600 Fourier transform infrared spectrometer (FTIR) equipped with a deuterated triglycine sulphate (DTGS) detector.

The microwave spectrum of the laser output was measured with a fast QWIP whose bandwidth is about 20 GHz. The laser output was focused directly onto the fast QWIP, and the resulting photocurrent was displayed on a spectrum analyzer. The resolution bandwidth of the spectrum analyzer was 330 Hz for the measurement.

The IAC measurement was based on a Michelson interferometer. The laser beam first passed through a chopper, and was then sent to a Ge beam splitter with antireflection coating on one side. The transmitted and reflected beams were then sent into two broadband retroreflectors coated with gold, with one of the retroreflectors mounted on a stepping motor. Once recombined by the beam splitter, the two beams were sent collinearly to a two-photon QWIP with operating wavelength centred at 6.2 μm. The resulting photocurrent was then sent to a current pre-amplifier, and subsequently to a lock-in amplifier whose reference frequency was determined by the chopper. The signal from the lock-in amplifier was recorded by a computer which controls the stepper motor simultaneously.

**Supplementary Information** accompanies the paper on **www.nature.com/nature**.

Acknowledgements: The authors thank W. F. Andress for support of network analyzer and useful discussions as well as Ariel Gordon and Christian Jirauschek for initial implementations of the simulation tools used. They also thank D. Bour, S. Corzine and G. Höfler at Agilent Technologies for wafer growth.

The authors declare no competing interests.

Correspondence and requests for materials should be addressed to F.C. (e-mail: capasso@seas.harvard.edu).

## Figures

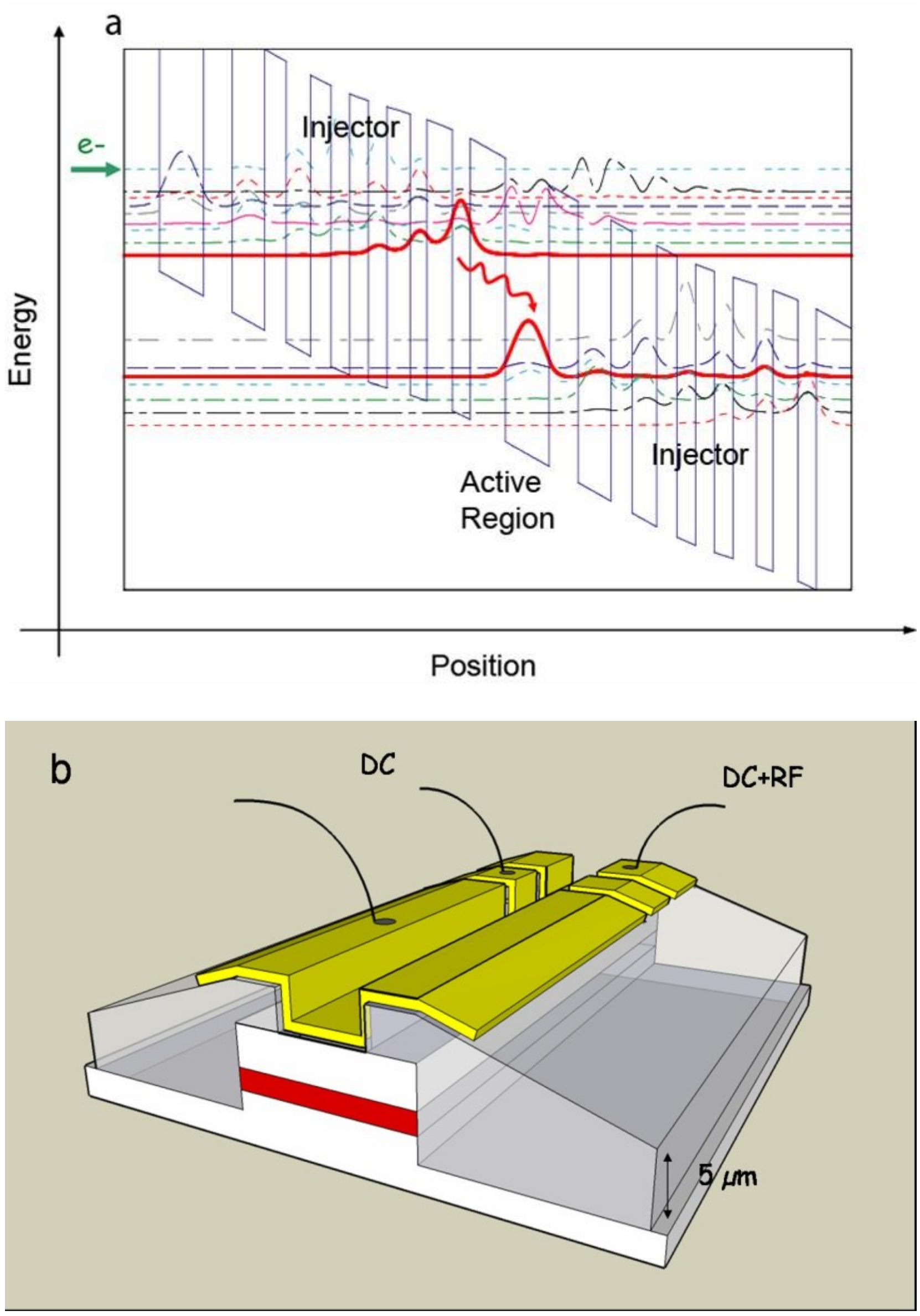

Figure 1. Diagrams showing the band structure design and the modulation scheme. (a) Calculated conduction band structure of one period of the QCL. The plot represents the potential profile along the growth direction, and the moduli squared of the wavefunctions. The states involved in the laser transition are shown as solid red curves. The barriers (conduction band offset = 0.52 eV) are made of $Al_{0.48}In_{0.52}As$ and the quantum wells of $Ga_{0.47}In_{0.53}As$. The layer thicknesses are (starting from the left, from the injection barrier, in nm):

**3.8**/4.7/**3.1**/3.5/**2.3**/2.6/**2.2**/2.0/**2.0**/2.0/**2.5**/1.8/**2.7**/1.9/**3.8**, where the barriers are indicated in boldface and the underlined layers are doped to n= 6x10$^{17}$ cm$^{-3}$. (b) Diagram of the multi-section QCL, showing the dry-etched laser ridges in white (with active region in red), the 5 μm-thick SU-8 insulating layer in grey, and the gold top contact in yellow. The top metal contact layer and the underlying heavily doped region grown above the top cladding layer are disconnected between the sections. The whole 2.6 mm-long laser is biased under the same DC voltage, while a RF modulation is added to the 120μm-long small section at the end of the ridge.

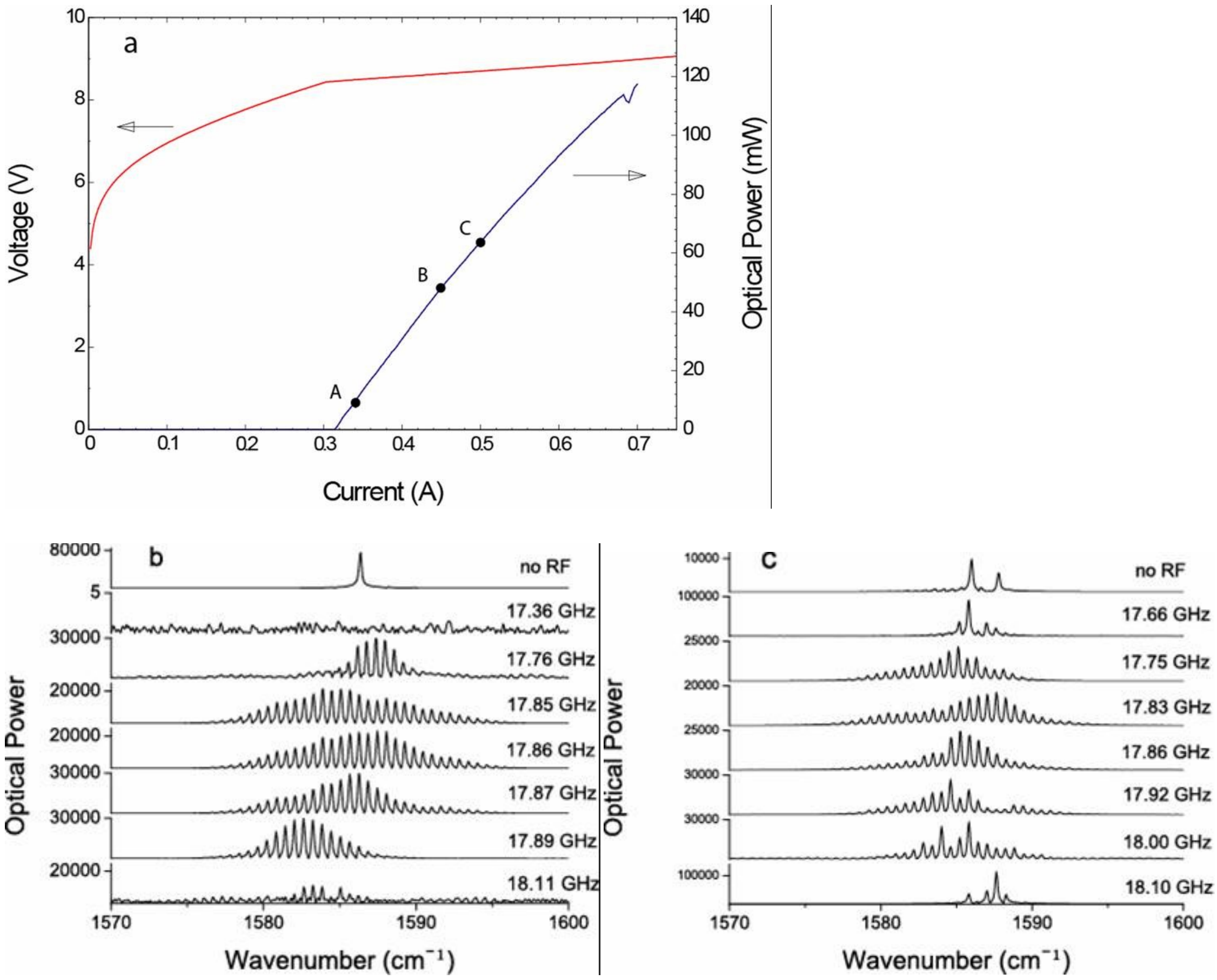

Figure 2. Laser characteristics. (a) The current-voltage (I-V) and output power-current (L-I) characteristics of the 16 μm wide QCL with no RF modulation. The laser threshold is at 310 mA. A, B, C denote the currents (I=340 mA, 450 mA &

500mA) at which currents the pulse characterization was performed (b) Spectra of the QCL at 340 mA and (c) at 450 mA with 35 dBm of input RF modulation at various modulation frequencies. Note that the scales of the optical power of the spectra are different at 340 mA. All measurements were performed at a heat sink temperature of T=77K.

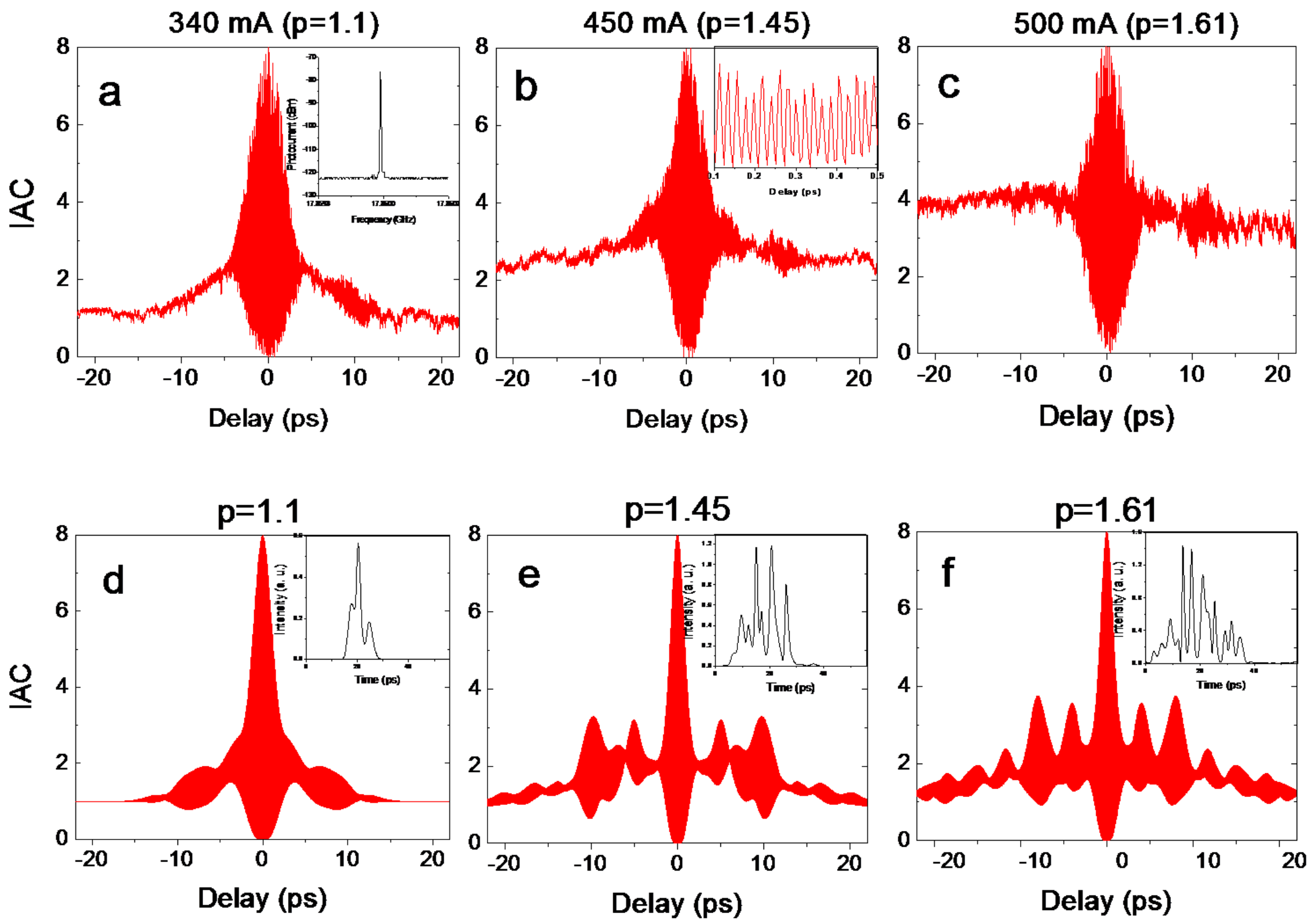

Figure 3. Pulse characterization of the 16 μm-wide QCL output with 35 dBm of applied modulation at various DC pumping levels and corresponding simulations. (a), (b), (c) Interferometric autocorrelation traces (IACs) of the QCL pumped at 340 mA (1.1 times the laser threshold), 450 mA (1.45 times the laser threshold) and 500 mA (1.61 times the laser threshold), respectively. Inset (a): Microwave spectrum of photocurrent generated by the laser at 340 mA. Inset (b): detail of the interference fringes from 0.1 ps to 0.5 ps. All measurements

were performed at a heat sink temperature of T=77K and RF modulation frequency at 17.86 GHz. (d), (e), (f) Simulated IACs at *p*=1.1, 1.45 & 1.61, respectively, with fixed modulation amplitude *m*=5. Insets show the corresponding intensity profile of one round-trip.

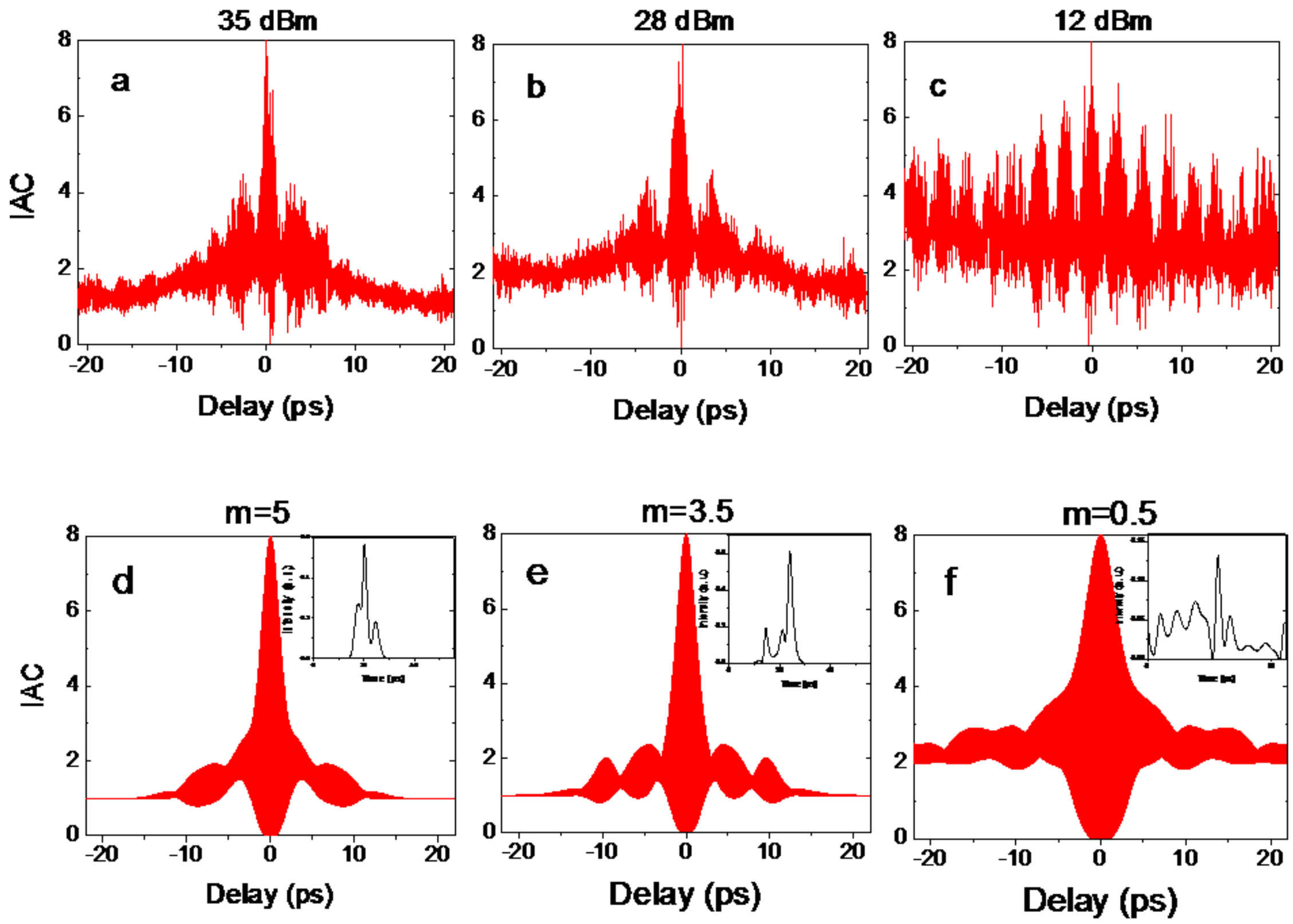

Figure 4. Pulse characterization of the 12 μm-wide QCL output at 265 mA with various RF input powers and corresponding simulations. (a), (b), (c) Interferometric autocorrelation traces (IACs) of the QCL with 35 dBm, 28 dBm & 12 dBm of RF input powers, respectively. All measurements were performed at a heat sink temperature of T=77 K and RF modulation frequency at 17.415 GHz. (d), (e), (f) Simulated IACs with modulation amplitude *m*=5, 3.5 & 0.5, respectively, with fixed DC pumping *p*=1.1. Insets show the corresponding intensity profile of one round-trip.